\documentclass[a4paper,fleqn,usenatbib]{mnras}

\usepackage{newtxtext,newtxmath}

\usepackage[T1]{fontenc}
\usepackage{ae,aecompl}

\usepackage{times}
\usepackage{amssymb}
\usepackage{amsmath}
\usepackage{color}
\usepackage{bm}
\usepackage{graphicx}
\usepackage{tabu}
\usepackage{epsfig}
\usepackage{longtable,tabu}
\usepackage{url}
\usepackage{xcolor}
\usepackage{ulem}
\usepackage{supertabular}
\usepackage{caption}
\usepackage[multidot]{grffile}
\usepackage{subcaption}
\DeclareSymbolFont{matha}{OML}{txmi}{m}{it}
\DeclareMathSymbol{\varv}{\mathord}{matha}{118}

\def\mnras{MNRAS}
\def\aj{AJ}
\def\aap{A\&A}
\def\apj{ApJ}
\def\apjl{ApJ}
\def\apjs{ApJS}

\def\pasp{PASP}
\def\pasj{PASJ}

\def\nat{Nature}

\def\lir{$L_{\rm IR}$}

\def\l1.4{$L_{\rm 1.4GHz}$} \def\s1.4{$S_{\rm 1.4GHz}$}

\def\xunits{M$_{\odot}$ ({\sc k}\,km\,s$^{-1}$ pc$^2$)$^{-1}$}
\def\kms{km\,s$^{-1}$}

\def\Nplus{[N\,{\sc ii}]}
\def\ha{{H$\alpha$}}

\def\jonezero{$J\!=\!1\!-\!0$}

\def\um{$\mu$m}
\def\targetname{HATLAS\,J084933}
\def\targetnamew{HATLAS\,J084933-W}
\def\targetnamee{HATLAS\,J084933-E}

\def\gs{\mathrel{\raise0.35ex\hbox{$\scriptstyle >$}\kern-0.6em
\lower0.40ex\hbox{{$\scriptstyle \sim$}}}}
\def\ls{\mathrel{\raise0.35ex\hbox{$\scriptstyle <$}\kern-0.6em
\lower0.40ex\hbox{{$\scriptstyle \sim$}}}}
\def\m@th{\mathsurround=0pt }
\def\eqalign#1{\null\,\vcenter{\openup1\jot \m@th
 \ialign{\strut\hfil$\displaystyle{##}$&$\displaystyle{{}##}$\hfil
 \crcr#1\crcr}}\,}

\title[Hyperluminous starburst gives up its secrets]
{Hyperluminous starburst gives up its secrets}

\author[Ivison et al.]{%
\Large 
R.\,J.~Ivison,\(^{\! 1,2}\) 
M.\,J.~Page,\(^{\! 3}\)
M.~Cirasuolo,\(^{\! 1}\) 
C.\,M.~Harrison,\(^{\! 1}\)
V.~Mainieri,\(^{\! 1}\)
V.~Arumugam\(^{1,4}\)
\and
\Large
and
U.~Dudzevi\v{c}i\={u}t\.e\(^{5}\)
\vspace*{1mm}\\
\(^1\) European Southern Observatory, Karl-Schwarzschild-Strasse~2, D-85748 Garching, Germany\\
\(^2\) Institute for Astronomy, University of Edinburgh, Blackford Hill, Edinburgh, EH9~3HJ\\
\(^3\) Mullard Space Science Laboratory, University College London, 
Holmbury St. Mary, Dorking, Surrey RH5~6NT\\
\(^4\) Institut de Radioastronomie Millim\'etrique, 300 rue de la Piscine, F-38406 Saint-Martin d'H\`eres, France\\
\(^5\) Centre for Extragalactic Astronomy, Durham University, South Road, Durham DH1~3LE
}

\date{
Accepted 2019 July 30. Received 2019 July 18; in original form 2019 May 24
}

\pubyear{2019}

\begin{document}
\label{firstpage}
\pagerange{\pageref{firstpage}--\pageref{lastpage}}
\maketitle


\begin{abstract}
  HATLAS\,J084933.4+021443 was identified as a dusty starburst
  via its rest-frame far-infrared (far-IR)
  emission. Multi-frequency imaging and spectroscopy revealed a
  cluster of four dusty galaxies at $z=2.41$, covering
  80\,kpc. Here, we use ALMA to confirm a more distant, fifth
  protocluster member, and present X-ray and rest-frame optical
  imaging spectroscopy of the brightest, an unlensed
  hyperluminous IR galaxy (HyLIRG). The data reveal broad \ha\
  and bright \Nplus\ lines, and bright X-ray emission,
  characteristics that betray a Type-1 active galactic nucleus
  (AGN), strengthening evidence that AGN are ubiquitous amongst
  HyLIRGs. The accreting black hole is super massive,
  $M_{\rm bh}\approx 2\times 10^9$\,M$_\odot$, with little
  intrinsic absorption,
  $N_{\rm H}\approx 5\times 10^{21}$\,cm$^{-2}$. The X-ray
  properties suggest the accretion luminosity rivals that of
  the starburst, yet it is not obvious where this
  emerges in its panchromatic spectral energy distribution (SED).
  We outline three scenarios that could give rise to the observed
  characteristics, and how we might distinguish between them.
  In the first, we see the AGN through the host galaxy because of
  the cavity it excavates.  In the others, the AGN is not
  co-spatial with the starburst, having been ejected via
  asymmetric gravitational radiation, or having evolved towards
  the naked quasar phase in an unseen companion.
\end{abstract}




\section{Introduction}
\label{introduction}

Since the discovery of the first example amongst thousands of faint
sources uncovered by the {\it Infrared Astronomical Satellite}, the
origin of the prodigious energy that characterises hyperluminous
infrared galaxies (HyLIRGs, defined such that \lir\
$\ge 10^{13}$\,L$_\odot$) has been a topic of controversy ---
`monsters or babies?' \citep{lutz99}.  Indeed, `hidden quasar or
protogalaxy?' was part of the title of the paper announcing the
discovery of the now-famous $z=2.3$ galaxy, IRAS\,FSC10214+4724
\citep{rr91}, which still qualifies as a HyLIRG even though it was
eventually found to be strongly lensed \citep{gl95, deane13}. A decade
later, this same rare population made an appearance amongst early
images obtained in the submm waveband, with the first submm-selected
galaxy (SMG) being identified as a weakly lensed HyLIRG at $z=2.8$,
comprising a dust-obscured, gas-rich starburst alongside a
broad-absorption-line (BAL) quasar \citep{ivison98leblob,
  ivison10leblob, frayer98, frayer18, vernet01}.

As a result of the spatial resolution of the Atacama Large Millimetre
Array (ALMA), many previously suspected HyLIRGs have been resolved
into multiple, discrete ultraluminous IR galaxies (ULIRGs), some
hovering around the HyLIRG threshold \citep[e.g.][cf.\
\citealt{younger07, younger08, riechers13}]{karim13, fu13,
  oteo16cplus, oteo18, riechers17, litke19}. Genuine, instrinsic
HyLIRGs are thus known to be extraordinarily rare --- indeed, to find
the nearest examples one must search out to $z\approx 0.3$
\citep[][cf.\ \citealt{efstathiou14}]{rrw10}.  Nevertheless, HyLIRGs
are excellent laboratories with which to confront recent hydrodynamic
simulations of isolated and merging galaxies \citep[e.g.][]{hayward11,
  narayanan15} which struggle to reproduce their number densities in
the presence of the feedback required to match the local mass
function. The luminosity of a HyLIRG implies a star-formation rate
(SFR) of $\approx 3,400$\,M$_\odot$\,yr$^{-1}$; indeed, its SFR would
remain substantial if the stellar initial mass function is top
heavy \citep{zhang18}, or if there is a substantial contribution to
\lir\ from a powerful AGN as has often been suspected
\citep[e.g.][]{hw93, franceschini00}. Either way, when observing
HyLIRGs we are witnessing galaxy formation at its most extreme and it
is important to understand which physical processes trigger these
far-IR-luminous events, and the subsequent quenching mechanisms. Are
these high-redshift starbursts similar to the ULIRGs seen locally,
with the same efficiency in converting gas into stars, or do they have
a higher star-formation efficiency? If the latter, why?  Are the
relations between metal content, star formation and mass similar to
other high-redshift galaxy populations?  How do the starburst episodes
relate to the growth of the central black holes?

As its name implies, HATLAS\,J084933.4+021443 (hereafter \targetname),
was found in the largest extragalactic {\it Herschel} survey, {\it
  H}-ATLAS \citep{eales10}, with $S_{\rm 350\mu m} = 249$\,mJy
\citep{valiante16}.  Its redshift was determined quickly via multiple
CO lines ($z=2.41$ -- \citealt{harris12, ivison13, gomez18}),
suggesting \lir\ $\approx 6\times 10^{13}$\,L$_\odot$.  Extensive
panchromatic observations, including imaging at high spatial
resolution with ALMA, the Jansky Very Large Array (JVLA), the
Submillimetre Array and the Institut Radioastronomie Millimetrique's
Plateau de Bure Interferometer (IRAM PdBI), revealed that \targetname\
-- like many other HyLIRGs \citep[e.g.][]{karim13} --- breaks down
into multiple gas-rich galaxies at the same redshift, covering
$\approx 10$\,arcsec or $\approx 80$\,kpc in the plane of the sky,
designated W, T, M and C (see\footnote{Alternatively, the layout of
  these galaxies is shown later in this paper.} Figure~1 of
\citealt{ivison13}\defcitealias{ivison13}{I13} hereafter
\citetalias{ivison13}), each component a distinct ULIRG or HyLIRG.  T
is gravitationally amplified\footnote{Weakly --- by less than a factor
  two; intrinsically, T is still a HyLIRG.} by a foreground
edge-on spiral; W is an unlensed HyLIRG. M and C are somewhat less
luminous (still \lir\ $>10^{12}$\,L$_\odot$, so ULIRGs, and unlensed)
yet gas-rich galaxies.

An unusually high intrinsic IR luminosity was first suspected for
\targetname\ because of its broad CO $J=1$--0 line,
$\approx 1,180$\,km\,s$^{-1}$ FWHM, as detected by the Greenbank
Telescope (GBT), which \citet{harris12} argued was consistent with no
gravitational amplification. However, although W does have an
unusually broad CO line, $825\pm 115$\,km\,s$^{-1}$ FWHM, and is
unlensed, the overall line width was shown by \citetalias{ivison13} to
owe much to the velocity dispersion of the aforementioned group of
luminous starbursts found within the GBT beam.

High-resolution ($\sim 0.3$-arcsec) 3-D spectroscopy obtained with
JVLA and the IRAM PdBI in $^{12}$CO $J=1$--0 and 4--3, respectively,
traced the molecular gas dynamics on scales of $\approx 1$\,kpc,
measuring the spatial extent of the gas ($\sim 1$\,arcsec, or
$\sim 8$\,kpc), its mass (and density), Toomre parameter and the
mid-plane hydrostatic ISM pressure. Later ALMA observations suggested
that the far-IR emission of the most luminous component, W,
corresponds to greybody emission from dust at a single temperature,
$\approx 40$\,K, throughout the full extent of the galaxy
\citep{gomez18}.

In all of these long-wavelength studies, and also with regard to
its rest-frame optical--through--radio SED, \targetnamew\
resembles a starburst, with no obvious sign of any influence from
an AGN, though --- like mergers and interactions --- accreting
black holes are often difficult to identify in dusty starbursts
with anything but the deepest and most complete of datasets.  In
this paper we present new observations of \targetname\ obtained
using the Atacama Large Millimetre/Submillimetre Array (ALMA),
the European Space Agency's {\it XMM-Newton} space observatory
and the KMOS spectrograph on UT1 of the European Southern
Observatory's Very Large Telescope, to further elucidate the
nature of this galaxy: hidden quasar, or protogalaxy?  Monster,
or baby?

The paper is organised as follows: \S\,\ref{sec:observations}
describes the observations and data reduction. \S\,\ref{sec:results}
presents the results, with our discussion of those results in
\S\,\ref{sec:discussion}.  We summarise our results and draw
conclusions in \S\,\ref{sec:summary}.  Throughout, we adopt a standard
$\Lambda$-CDM cosmology with $\Omega_{\rmn m} = 0.3$,
$\Omega_\Lambda = 0.7$ and $H_0 = 70$\,km\,s$^{-1}$\,Mpc$^{-1}$, such
that 1\,arcsec corresponds to 8.1\,kpc at $z=2.41$.

\section{Observations and data reduction}
\label{sec:observations}

\subsection{KMOS/VLT}

The KMOS observations were carried out during ESO observing period
P94, under the observing programme $\rm 094.A-0214(A)$.  We adopted
the standard object-sky-object nod-to-sky observation pattern, with
300-s exposures and an alternating 0.2- and 0.1-arcsec dither pattern
to improve the spatial sampling. The target was observed in the $K$
band, with a total integration time (on source) of 80\,min, with a
median seeing of 0.7\,arcsec.

The data reduction was performed by using SPARK \citep[Software
Package for Astronomical Reduction with KMOS --][]{davies13},
implemented using ESOREX \citep[ESO Recipe Execution Tool
--][]{freudling13}.  Each of the 300-s exposures was re-constructed
independently, wavelength calibrated and sky subtracted using the
closest sky exposure, and finally re-sampled into a data cube with
$0.1\times 0.1$\,arcsec$^2$ spaxels. In order to improve the sky
subtraction we used the {\sc skytweak} option within SPARK
\citep{davies07}, which accounts for the time variability of the
various OH sky-line families. Standard-star observations were carried
out on the same night as the science observations and used for
telluric correction and flux calibration. The individual 300-s cubes
were finally combined together to create a stacked cube.

\subsection{XMM-Newton}

\begin{figure*}
\centering  
\includegraphics[width=5.1in, angle=90]{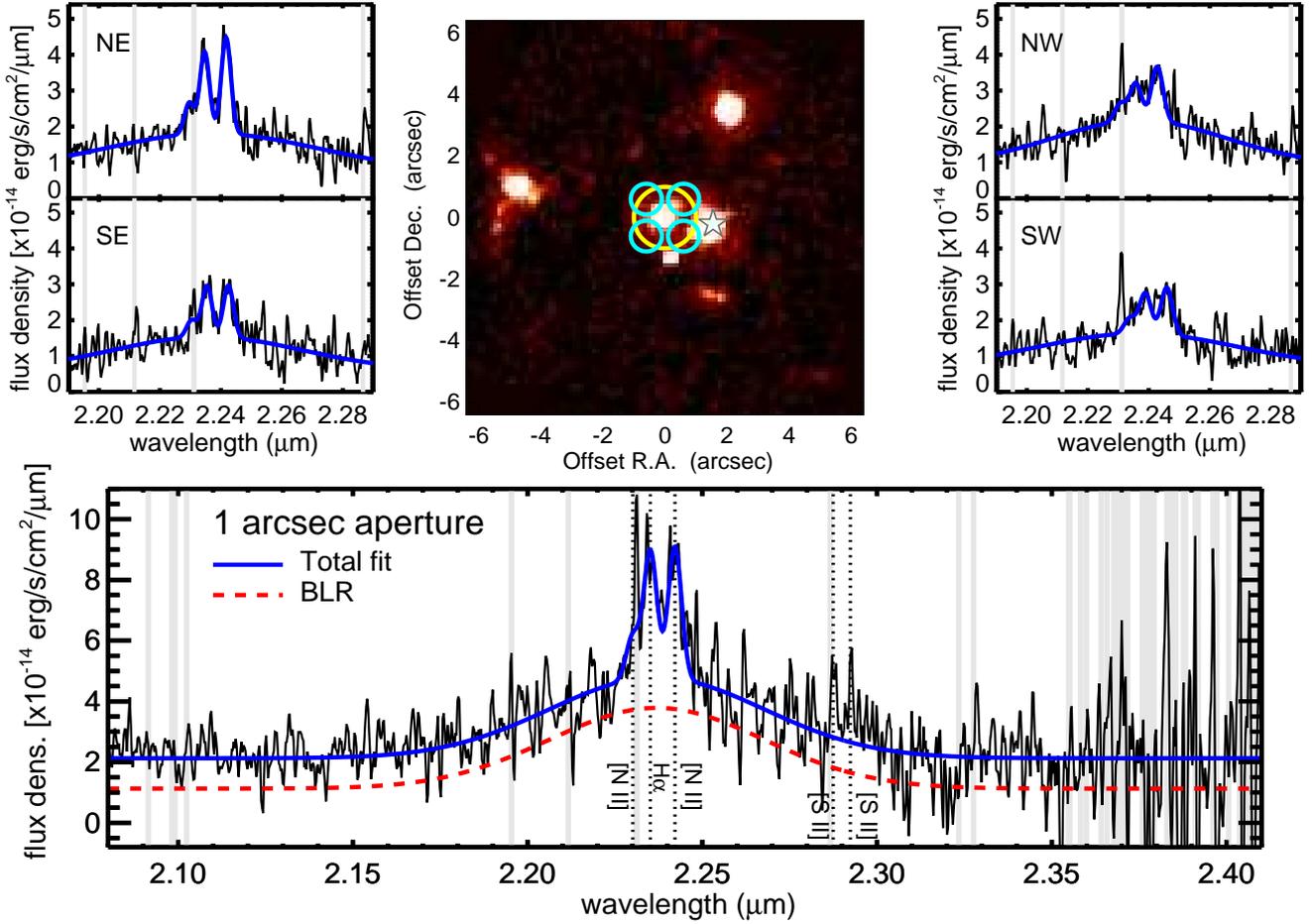}
\caption{Extracted inside a 1-arcsec-diameter aperture (yellow circle,
  overlaid on the {\it Hubble Space Telescope (HST)} F110W image from
  \citetalias{ivison13}), the lower panel shows the broad H$\alpha$
  emission, spanning 9,700\,km\,s$^{-1}$ FWHM, in component W of
  \targetname\, as observed by VLT/KMOS, unambiguously revealing the
  presence of a previously unseen broad-line AGN. [N\,{\sc ii}] at
  654.986, 658.527\,nm and [S\,{\sc ii}] can also be seen (and is also
  marked).  The other four spectra illustrate the variation in the
  ratio of [N\,{\sc ii}] to H$\alpha$, which increases markedly to the
  North-East (NE), as sampled by small off-centre apertures (the four
  blue circles).  It is unfortunate that the blue [N\,{\sc ii}] line
  falls almost exactly on a sky line (this region was masked during
  the fitting); however, this issue is mitigated by the fact that this
  blue [N\,{\sc ii}] line has a fixed wavelength offset and flux ratio
  ($3.06^{-1}\times$) with respect to the redder (brighter) component
  of the [N\,{\sc ii}] doublet. Given the bright red component, we can
  be confident that the blue component is buried under the sky line. N
  is up; E is to the left; offsets from
  $\alpha_{2000} = 132^{\circ}.3900$ and
  $\delta_{2000} = 2^{\circ}.2457$ are marked in arcsec. A known star
  is labelled.}
\label{fig:kmos}
\end{figure*}

\targetname\ was observed with {\it XMM-Newton} on 2017 April 22--23
(AO15, proposal 078435). The resulting data from the European Photon
Imaging Camera (EPIC), which comprises three X-ray charge-coupled device
cameras operated in a so-called `photon-counting mode', were reduced
with version 15.0 of the {\it XMM-Newton} Science Analysis Software
(SAS). The {\it XMM-Newton} observation was significantly affected by
particle background flaring events, from which the data were screened
on the basis of the full-field count rate above 5\,keV. After
filtering out the high-background periods, the net exposure times were
27, 26 and 26\,ks in the MOS\,1 and MOS\,2 cameras and the pn camera,
respectively.

Images were constructed in four bands: 0.2--0.5, 0.5--2, 2--5 and
5-10\,keV. Background images were constructed following the procedure
described by \citet{loaring05}. The images were searched
simultaneously for sources in the four energy bands using the standard
SAS tasks, {\sc eboxdetect} and {\sc emldetect}. The astrometric
solution of the images was refined by cross-correlating the X-ray
source list with optical sources from the Sloan Digital Sky Survey
\citep[SDSS --][]{adelman08} using the SAS task, {\sc eposcorr}.

A point-like X-ray source was found at
$\alpha = \rm 08h\,49m\,33.59s$,
$\delta = +02^\circ\,14'\,44.''6$ (J2000) with a 1-$\sigma$
statistical uncertainty of 0.4\,arcsec, which includes the
contribution from the astrometric cross match to SDSS.  This
position is coincident with that of the unlensed\footnote{An
  optical point source visible immediately to the west of W is a
  star.  As described by \citetalias{ivison13}, it is unable to
  provide any significant gravitational amplification; the X-ray
  data reported here are not compatible with stellar emission,
  thus we can reliably associate the X-ray emission with W.}
component, W, of \targetname. We have verified by examination of
the 3XMM catalogue \citep{rosen16} that our 0.4-arcsec positional
uncertainty is reasonable for an on-axis source with comparable
signal-to-noise ratio.  Note that the point spread function (PSF)
of EPIC, at around 6\,arcsec full-width half maximum (FWHM), is
sufficiently large that minor contributions to the X-ray flux
from the other components of the \targetname\ system may be
hidden in the wings of the PSF. Nevertheless, the precise
positional coincidence of the X-ray source with component W
implies that the X-ray emission is dominated by this component.

We extracted X-ray spectra in the three EPIC cameras from a
circular region of 15~arcsec radius, centred on the X-ray
source. Background spectra were obtained from an annular region
surrounding the source, from which detected X-ray sources were
excised.  Event patterns 0--12 were included in the spectra
derived from MOS, while for the spectra derived from the pn
camera we used patterns 0--4 above 0.4\,keV and only pattern~0
between 0.2 and 0.4\,keV. For MOS, channels corresponding to the
strong 1.5-keV Al K$\alpha$ background emission line
\citep{lumb02} were excluded.  The spectra of the target from the
different EPIC cameras were then combined to form a single
spectrum, and the corresponding response matrices and background
spectra were combined in an appropriate fashion to form a single
response matrix and a single background spectrum, following the
method described in Appendix~A of \citet{page03}. Finally, the
spectrum was grouped to a minimum of 20 counts per bin.

\subsection{ALMA}

The ALMA band-6 245-GHz (1.22-mm) data used here were obtained for
project 2013.1.00164.S, targeting the CH$^+$ line
\citep{falgarone17}. Our resulting continuum image, where we have
discarded the channels around the CH$^+$ line, was made using the {\sc
  clean} task in the Common Astronomy Software Application package
\citep[CASA --][]{mcmullin07}.  The image has an r.m.s.\ noise level,
$\sigma= 38\,\mu$Jy\,beam$^{-1}$, and the synthesised beam measures
$0.49 \times 0.48$\,arcsec$^2$ FWHM, with the major axis at a position
angle, measured East of North, of $127^{\circ}$.

\section{Results}
\label{sec:results}

Here, we outline what can be deduced from the new rest-frame
optical spectroscopy, X-ray spectroscopy, and the submm imaging,
as described in the previous section.

Fig.~\ref{fig:kmos} presents our new KMOS spectrum of component W of
\targetname, extracted in a 1-arcsec-diameter aperture.  The Balmer
H$\alpha$ 656.461-nm emission is very strong indeed, and broad, where
in SMGs it is normally a combination of weak and narrow lines offset
spatially from broader (few $\times 1000$\,\kms) compact components
\citep{md13}.  We fit Gaussians simultaneously to the broad and narrow
H$\alpha$ components, and to the [N\,{\sc ii}] doublet at 654.986 and
658.527\,nm with a red/blue [N\,{\sc ii}] line ratio of 3.06, with all
the narrow lines tied to the same redshift and line width\footnote{[S\,{\sc
  ii}] can also be seen, weakly; this wavelength range was excluded
from the fits.}, and we also fit simultaneously to the continuum. The
strongest sky lines (marked in grey in Fig.~\ref{fig:kmos}) were
masked during the fitting process. We find that the broad [narrow]
H$\alpha$ lines span 9,700 [600]\,km\,s$^{-1}$ FWHM, with the narrow
lines\footnote{Recall that the three CO lines observed towards
  \targetnamew\ have an error-weighted average redshift,
  $z_{\rm lsr}=2.4068\pm 0.0002$, so offset away from us along the
  line of sight from the rest-frame optical lines by
  $\approx 600$\,km\,s$^{-1}$.} at $z_{\rm lsr}=2.4048$. The line
fluxes determined for the narrow and broad H$\alpha$ lines, accurate
to $\approx 10$ per cent, are $2.0\times 10^{-17}$ and
$2.0\times 10^{-16}$\,ergs\,s$^{-1}$\,cm$^{-2}$.  The flux and width
of the broad line imply a black hole mass,
$M_{\rm bh} \approx 2\times 10^9$\,M$_\odot$ \citep{schulze18}, i.e.\
a super-massive black hole (SMBH).

There can be no doubt, therefore, based on the characteristics of
the H$\alpha$ emission line profile, that a broad-line Type-1 AGN
is present in the \targetnamew\ system.

In passing we note that the ratio of [N\,{\sc ii}] to H$\alpha$
increases markedly to the NE of the combined continuum/line centroid,
as illustrated in Fig.~\ref{fig:kmos}, suggestive of an ionisation
cone.

\begin{figure}
\includegraphics[width=2.42in,angle=-90]{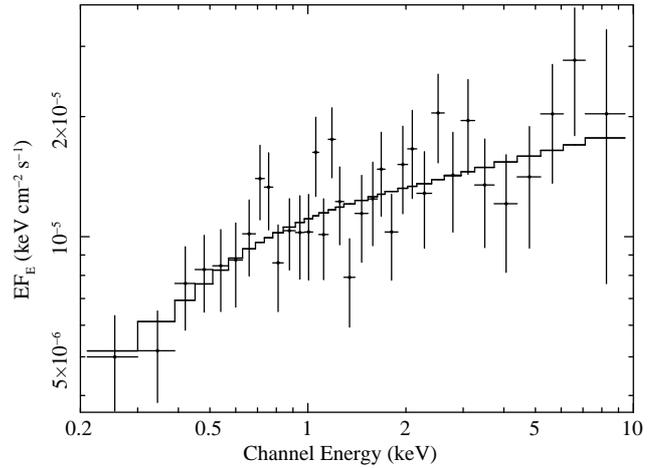}
\caption{\,The {\it XMM-Newton} EPIC X-ray spectrum of
  \targetname. The data points correspond to the measured
  spectrum and the stepped line corresponds to the best-fitting
  absorbed power-law model. Both model and data have been divided
  by the product of the effective area and Galactic column as a
  function of energy and are plotted as $EF_{\rm E}$, channel
  energy multipled by the energy flux per unit energy, so that an
  unabsorbed power law with $\alpha = +1$ would correspond to a
  horizontal line.}
\label{fig:xrayspectrum}
\end{figure}

\begin{figure}
\includegraphics[width=3.3in,angle=0]{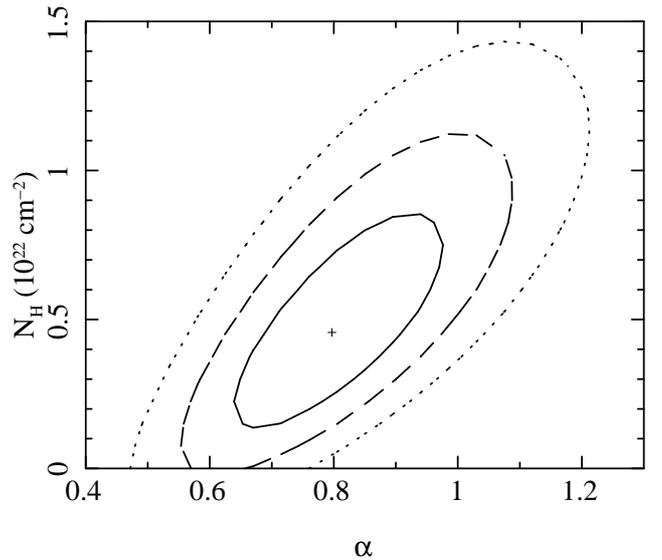}
\caption{Confidence contours for power-law slope, $\alpha$, and
  intrinsic column density, $N_{\rm H}$, from the model fit to
  the {\it XMM-Newton} EPIC X-ray spectrum. The solid, dashed and
  dotted contours correspond to $\Delta \chi^{2}$ of 2.3, 6.2 and
  11.8 respectively, equivalent to 68, 95 and 99.7 per cent
  confidence regions for two interesting parameters. The best-fit
  model parameters are indicated with a small cross.}
\label{fig:xraycontours}
\end{figure}

We turn now to the data from {\it XMM/Newton}. The X-ray spectrum
was modelled using version 11.3 of the X-ray spectral fitting
package, XSPEC. An absorbed power-law model was fitted to the
spectrum, in which the absorption was the product of cold
Galactic photoelectric absorption with a fixed column density of
$N_{\rm H} = 2.9\times 10^{20}$\,cm$^{-2}$, and cold
photoelectric absorption at $z=2.41$, corresponding to the rest
frame of \targetname, for which the column density was a free
parameter in the fit. The spectrum and best-fitting model are
shown in Fig.~\ref{fig:xrayspectrum}.  The fit yielded a
$\chi^{2}$ of 22.3 for 31 degrees of freedom, implying that the
model fits the data well.  Fig.~\ref{fig:xraycontours} shows the
confidence contours for the intrinsic ($z=2.41$) column density,
$N_{\rm H}$, and power-law energy index, $\alpha$, defined such
that $S_{\nu}\propto \nu^{-\alpha}$. The best-fitting power-law
index, $\alpha = 0.8\pm 0.1$, typical for QSOs at $z\approx 2$--3
\citep[e.g.][]{mateos05, mateos10}. The best-fitting intrinsic
column density, $N_{\rm H} =(5\pm 3) \times
10^{21}$\,cm$^{-2}$. Zero intrinsic column density corresponds to
a $\Delta \chi^{2}$ of 5.9, so the intrinsic absorption is only
marginally significant ($2\sigma$). The $3\sigma$ upper limit to
the intrinsic column is $1.2\times 10^{22}$\,cm$^{-2}$. The
best-fitting model implies a 2--10-keV flux of
$4.1\times 10^{-14}$\,erg\,cm$^{-2}$\,s$^{-1}$ and a 2--10-keV
luminosity, $L_{\rm X}=1.4\times 10^{45}$\,erg\,s$^{-1}$, once
corrected for Galactic and intrinsic absorption.

This X-ray luminosity is orders of magnitude brighter\footnote{We must
  acknowledge, of course, that we cannot yet know whether the X-ray
  emission is variable.} than we would expect for X-ray emission due
to star formation \citep[$L_{\rm X}\ls 0.004$\,\lir\
--][]{alexander05a} and brighter than any of the 21 X-ray-luminous
SMGs found by \citet{stach19} amongst the 274 SMGs covered by the
$\ge200$-ks {\it Chandra} X-UDS \citep{kocevski18} observations. We
find $L_{\rm X} \sim 0.011$\,\lir, roughly consistent with both the
`AGN-classified SMGs' of \citet{alexander05b} and the quasars
catalogued by \citet{elvis94}, with both $L_{\rm X}$ and \lir\ similar
to IRAS\,F15307+3252, a hyperluminous Seyfert 2 quasar at $z=0.93$
about which relatively little is known \citep{cutri94, ruiz07}.

Our X-ray data thus corroborate the conclusion of our rest-frame
optical spectroscopy: \targetnamew\ hosts an AGN.

To estimate the {\it bolometric} luminosity of the AGN, we begin by
translating the $K$-corrected X-ray luminosity to an expected
rest-frame ultraviolet (UV) 250-nm luminosity via the logarithmic
slope, $\alpha_{\rm OX}$, which connects the 250-nm and 2-keV points
on the SED. Using equation 4 from \citet{just07}, we
obtain\footnote{Note that the minus sign is explicit in our definition
  of $\alpha_{\rm OX}$ but not in that used by \citet{just07}.}
$\alpha_{\rm OX}=1.63$. Taking the resulting UV luminosity, we use the
bolometric correction from figure~12 of \cite{richards06} to arrive at
an overall bolometric correction factor of $110\times$ the 2--10-keV
luminosity, implying a bolometric luminosity of
$1.6\times 10^{47}$\,erg\,s$^{-1}$, or log
$L_{\rm bol}/{\rm L}_\odot\approx 13.62$, so of the same order as the
IR luminosity of the starburst, log
$L_{\rm IR}/{\rm L}_\odot=13.52\pm 0.04$ \citepalias{ivison13}.

To assess the uncertainty in our estimate of the AGN's bolometric
luminosity, we combine in quadrature the r.m.s.\ scatter in
$\alpha_{\rm OX}$ from \citet{strateva05} and in the UV-to-bolometric
correction factor from \citet{richards06}, obtaining an overall
$1\sigma$ logarithmic uncertainty of 0.31, so roughly a factor of
two.

In the absence of a measurement of the Balmer decrement, we can make a
crude estimate of the optical extinction towards the AGN using the
well-documented correlation between the logarithms of H$\alpha$ and
X-ray luminosity. For this, we use the regression of the full sample
of \citet{panessa06}, given by the fourth row of their table~3 and
shown in the right-hand panel of their figure~4, adjusting for the
different value of $H_{0}$ that they adopted. For our observed
2--10-keV luminosity, we would expect an intrinsic H$\alpha$
luminosity of $4.7\times 10^{43}$\,erg\,s$^{-1}$, where the measured
flux of the broad H$\alpha$ component corresponds to a luminosity of
$9\times 10^{42}$~\,erg\,s$^{-1}$. To reconcile these predicted and
observed luminosities requires 1.8~magnitudes of extinction. Taking
the scatter of the \citet{panessa06} sample about the regression into
account, together with the measurement errors on the H$\alpha$ and
X-ray luminosities, we estimate a $1\sigma$ uncertainty of
1.3~magnitudes on the implied extinction. Translating to $A_V$ from
the extinction experienced at the wavelength of H$\alpha$ using the
$R_{V}=3.1$ extinction law of \citet{cardelli89} gives
$A_{V}=2.2\pm1.6$.  \citet{guver09} suggest that
$N_{\rm H}=(2.21\pm 0.09)\times 10^{21} A_{V}$ for a Milky-Way-like
dust-to-gas ratio, so we find
$N_{\rm H}=(4.8\pm 3.5)\times 10^{21}$\,cm$^{-2}$, perfectly
consistent with the column density we determined using
our X-ray measurements.

Any intrinsic absorption towards the AGN is thus likely to be
rather small, which is perhaps surprising for a galaxy that
contains such immense quantities of molecular gas, and with
$\approx 2\times 10^9$\,M$_\odot$ of dust that we have already
noted is well fit with greybody emission at a single temperature
over the full extent of the galaxy \citep{gomez18}, bearing in
mind that typical SMGs are thought to be optically thick out to
$\approx 75$\,$\mu$m, with $N_{\rm H}\approx 10^{24}$\,cm$^{-2}$
and $A_{V}\approx 500$ \citep{simpson17}.
 
\begin{figure}
\centering 
\includegraphics[width=3.3in]{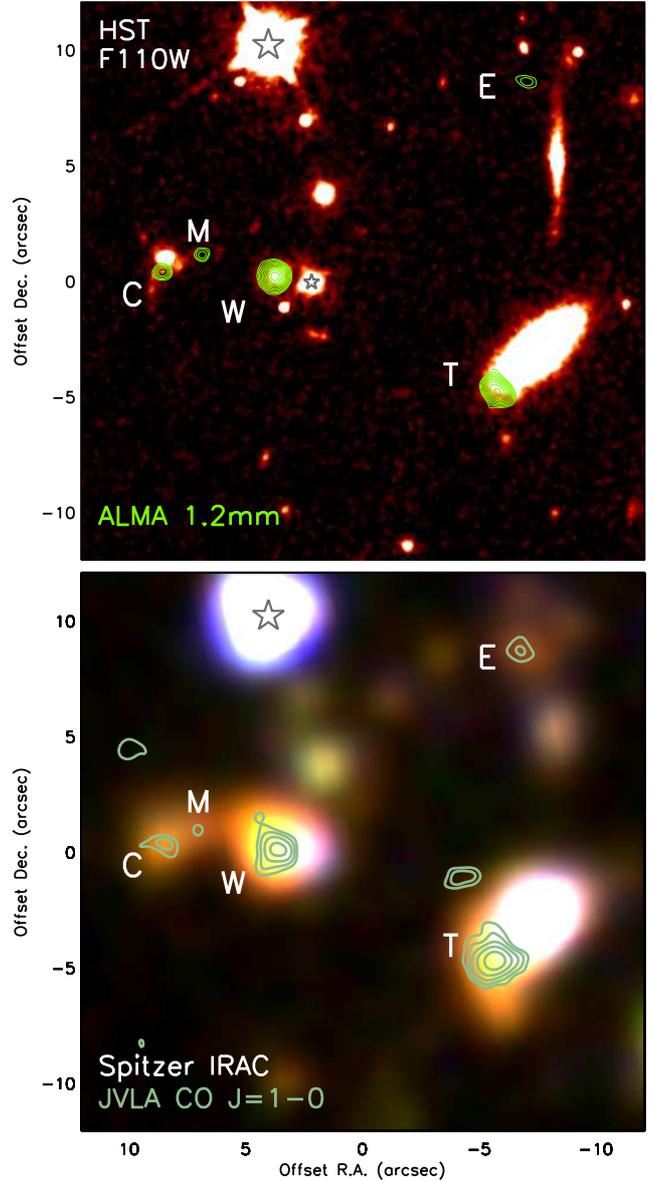}
\caption{{\it Top:} band-6 245-GHz (1.22-mm) continuum emission
  superimposed as contours on the {\it HST} F110W imaging of
  \targetname.  {\it Bottom:} contours of $^{12}$CO $J=1$--0 emission,
  collapsed optimally for each object, superimposed on a three-colour
  image comprising a heavily smoothed $J + H + K_{\rm s}$ image from
  VISTA as the blue channel plus the {\it Spitzer} IRAC 3.6- and
  4.5-$\mu$m data, all from \citetalias{ivison13}.  The object marked
  `E', north of T, can now be considered robustly identified --- via
  the positional coincidence of faint 1.2-mm and CO emission with a
  red IRAC counterpart --- as a fifth dusty, gas-rich galaxy in this
  $z=2.4$ proto-cluster. Contours are plotted at $-3, 3 \times \sigma$
  , with $\sqrt 2$-spaced increments thereafter, where $\sigma$ is the
  local noise level. N is up; E is to the left; offsets from
  $\alpha_{2000} = 132^{\circ}.3889$ and
  $\delta_{2000} = 2^{\circ}.2457$ are marked in arcsec. Known stars
  are labelled.}
\label{fig:vinod}
\end{figure}

Moving now to the ALMA data, Fig.~\ref{fig:vinod} presents our ALMA
band-6 continuum image of \targetname\, with contours superimposed on
the {\it HST} F110W image from \citetalias{ivison13}. In a separate
panel of Fig.~\ref{fig:vinod} we show the JVLA $^{12}$CO $J=1$--0
imaging and the {\it Spitzer} IRAC imaging from \citetalias{ivison13}.
We do this to illustrate that a band-6 continuum source, detected at
$6\sigma$, which we denote component `E', which lies
$\approx 12$\,arcsec to the north of component T, can now be seen to
be coincident with a very red galaxy that was detected by IRAC, as
well as being detected in CO, also at $6\sigma$, but which is not seen
in the F110W {\it HST} image.  E lies in a confused region of the maps
obtained by the {\it Wide-field Infrared Survey Explorer} \citep[{\it
  WISE} --][]{wright10} and a limit cannot be set that is both
meaningful and useful. Component E can thus be added to the inventory
of dusty, gas-rich galaxies that form part of the protocluster
associated with \targetname, making five in total, covering
$\approx 15$\,arcsec or $\approx 120$\,kpc.

As seen in the CO $J=1$--0 image, integrated over its full line width,
E subtends $\approx 7.3$\,kpc.  Consistent measurements of
$I_{\rm CO}$ were obtained from this image and from a Gaussian fit to
the spectrum extracted at the peak and corrected for
$I_{\rm total}/I_{\rm peak}$. E lies close to the redshifts of M and
C, somewhat redward of W and T, and thus further helps to explain the
broad line seen originally by the $\approx 22$-arcsec primary beam
of GBT \citep{harris12}.  The line is considerably broader than those
of the other cluster galaxies, $\approx 1,380$\,km\,s$^{-1}$ FWHM, albeit
with a large uncertainty. The signal to noise is too low for us to be fully
confident, but two gas clumps may be involved, distinct both spatially
and in redshift, or there could be a rotating gas disk, as with W and
T, this time along PA $\approx 45^\circ$, with the reddest component
to the north east.  Scaling from the band-6 flux density and SED of
component W, component E has an IR luminosity of approximately
$6\times 10^{12}$\,L$_\odot$.  Following \citet{kennicutt98}, this
implies an SFR of 650\,M$_\odot$\,yr$^{-1}$ for a \citet{chabrier03}
stellar initial mass function (IMF), or considerably less for the IMF
observed in distant, dusty starbursts by \citet{zhang18}.  E contains
approximately $6\times10^{10}$ and $2\times10^{10}$\,M$_\odot$ of gas
and stars, respectively.  Its basic observational properties are
listed in Table~\ref{tab:e}.

\begin{table}
{\centering
\caption{Basic observational properties of component~E \label{tab:e}}
\begin{tabular}{rcl} \\
  Wavelength & $S_{\nu}$ & Comment\\
\hline
1.1 \um & $3\sigma<0.8$&$\mu$Jy; F110W\\
2.15 \um & $3\sigma<9.1$&$\mu$Jy; VISTA $K_{\rm S}$\\
3.6 \um & $7.9\pm 1.0$ &$\mu$Jy; IRAC\\
4.5 \um & $10.3\pm 1.5$ &$\mu$Jy; IRAC\\
1.22 mm$^{\rm a}$ & $1.47\pm 0.24$&mJy; ALMA\\
5.9 cm  & $3\sigma<45$ &$\mu$Jy; JVLA\\
\hline
R.A.\ & 08:49:32.867 ($\pm 0.003$) & J2000\\ 
Dec.\ &+02:14:53.12 ($\pm 0.03$)   &J2000\\ 
log \lir$^{\rm b}$ &$12.8^{+0.1}_{-0.2}$&L$_{\odot}$\\
SFR & 650 & M$_\odot$\,yr$^{-1}$ (see text)\\
FWHM CO \jonezero\ & $7.3\pm 1.8$ &kpc\\ 
CO \jonezero\ $I_{\rm CO}$ &  $0.27\pm 0.06$ & Jy\,\kms \\
CO \jonezero\ FWHM& $1380\pm 410$ &km\,s$^{-1}$\\ 
CO \jonezero\ $z_{\rm LSR}$& $2.4151\pm 0.0018$&\\  
CO \jonezero\ $L^\prime_{\rm CO}$ &  $76\pm 17$ & $10^9$\,{\sc k}\,\kms\,pc$^2$ \\  
CO \jonezero\ $L_{\rm CO}$ &  $3.9\pm 0.9$ &$10^6$\,L$_\odot$\\  
log $M_{\rm H_2+He}^{\rm c}$ &$10.8\pm 0.2$&M$_{\odot}$\\
log $M_{\rm stars}$ &$10.3\pm 0.2$&M$_{\odot}$\\
SFE &120&L$_{\odot}$\,M$_{\odot}^{-1}$\\
\hline 
\end{tabular}}

\noindent
$^{\rm a}$Peak 1.22-mm flux density, $660\pm 75\,\mu$Jy\,beam$^{-1}$, so resolved
by the $0.49 \times 0.48$\,arcsec$^2$ FWHM beam.

\noindent
$^{\rm b}$Scaled from component W, where
$S_{\rm 1.22mm}=8.29\pm 0.13$\,mJy and log \lir\ = 13.52, thereby
adopting the same SED as W for component E.

\noindent
$^{\rm c}$For $\alpha_{\rm  CO}=0.8$\,\xunits.
\end{table}

Component E is not detected in our {\it XMM-Newton} images, and
is in the wings of the PSF of component W.  We measured the
counts in a 10-arcsec-radius aperture at the position of
component E and can set a $3\sigma$ 2--10-keV upper limit of
$8.5\times 10^{-15}$\,erg\,cm$^{-2}$\,s$^{-1}$.

\section{Discussion}
\label{sec:discussion}

Before discussing plausible explanations for the properties of
\targetnamew, let us first look at our results in the context of
other relevant samples.  The X-ray properties of \targetnamew\
are consistent with those of the majority of the 14 HyLIRGs at
$z=0.3$--2.0 observed in X-rays by \citet{ruiz07}, where the
sample was selected to contain a range of examples of the HyLIRG
population, including Type 1 and 2 QSOs, so significantly biased
towards those containing AGN. Compared to the \citet{lusso12}
sample of 929 AGN, selected over a $\approx 2$-deg$^2$ field via
X-rays, only five have higher bolometric luminosities than
\targetnamew; its redshift is amongst the most distant decile of
Type-1 AGN; its absorbing column is roughly $7\times$ the average
$N_{\rm H}$ for an X-ray-selected Type-1 AGN --- not particularly
unusual --- and half the average for a Type-2 AGN.

We know that there are no clear signatures of an AGN in the
rest-frame UV spectrum of \targetnamew, where the Keck
spectroscopy of \citetalias{ivison13} revealed only faint C\,{\sc
  ii}] at rest-frame 232.6\,nm, consistent with considerable dust
obscuration.  Is there any other indication from existing
observations of \targetnamew\ that it might harbour a powerful
AGN?  Is the panchromatic SED of \targetnamew\ more consistent
with a typical SMG, or with a luminous AGN?

The SED of \targetnamew\ was not presented blueward of
$\lambda_{\rm obs}=880$\,nm by \citetalias{ivison13}, but it has since
been observed
as part of the wide layer of the Hyper Suprime-Cam Subaru Strategic
Program \citep{aihara18}, with $g=23.69\pm 0.26$, $r=22.59\pm 0.18$,
$i=21.92\pm 0.13$, $z=21.41\pm 0.09$ and $y=21.08\pm 0.08$\,AB~mag.
We also note that {\it WISE} obtained a $3\sigma$ detection consistent
with the position of component W of \targetname\ - albeit with poor
spatial resolution, 6.5\,arcsec FWHM --- in its W3 band at 12\,$\mu$m,
with a flux density of $420\pm140$\,$\mu$Jy.  An even more tentative
($<3\sigma$) detection was made in the W4 band (12\,arcsec FWHM) at
22\,$\mu$m: $3.3\pm 1.3$\,mJy. These data represent weak evidence that
the mid-IR SED tends towards the AGN region of colour-colour space
outlined in diagnostic plots \citep[e.g.][]{ivison04, lacy04}.

\begin{figure}
\centering 
\includegraphics[width=3.3in]{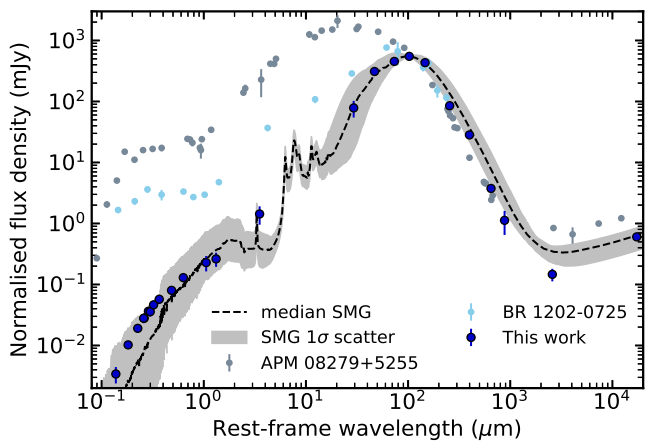}
\caption{Rest-frame ultraviolet--through--radio SED of  
  \targetnamew, with new photometry from the Hyper Suprime-Cam  
  Subaru Strategic Program \citep{aihara18} and from {\it WISE}.  
  Also shown are the median SED of ALMA-identified SMGs [dashed  
  line] in the Ultra-Deep Survey field (U.~Dudzevi\v{c}i\={u}t\.e et  
  al., in prep), where the grey area is the r.m.s.\ spread in SMG  
  SEDs, and the SEDs of the Type-1 dust-rich quasars,  
  APM\,08279+5255, at $z=3.9$ \citep{irwin98, leung19} and  
  BR\,1202$-$0725, at $z=4.7$ \citep{mcmahon94,leipski10}.  The  
  SEDs have been normalised at rest-frame 100\,$\mu$m.  Across  
  rest-frame $\approx  
  0.15$--30\,$\mu$m, an order of magnitude or more in flux density  
  separates the SEDs of \targetnamew\ and the dusty Type-1 AGNs.}
\label{fig:sed}
\end{figure}

An updated SED for \targetnamew, now spanning rest-frame
UV--radio wavelengths, is shown in Fig.~\ref{fig:sed}.  For
comparison, Fig.~\ref{fig:sed} also shows the median SED of over
700 ALMA-identified SMGs from U.~Dudzevi\v{c}i\={u}t\.e et al.\
(in prep), and the well-sampled SEDs of the dust-rich, Type-1
AGNs, APM\,08279+5255 and BR\,1202$-$0725 \citep{irwin98,
  mcmahon94, leipski10, leung19}. The SEDs have been normalised
at rest-frame 100\,$\mu$m.

Compared to a typical SMG, Fig.~\ref{fig:sed} reveals that an
extra $(2.8\pm0.9)\times 10^{11}$\,L$_\odot$ emerges from
\targetnamew\ across the rest-frame UV-optical wavelength regime,
where it is a magnitude bluer in $g-K$.  The SED of \targetnamew\
is, however, fully consistent with that of an SMG, lying at the
upper boundary of the r.m.s.\ scatter in rest-frame UV--optical
emission from typical SMGs, whereas we can see that an order of
magnitude or more in flux density separates the SEDs of
\targetnamew\ and the dusty Type-1 AGNs, across two orders of
magnitude in wavelength,
$\lambda_{\rm rest}\approx 0.15$--30\,$\mu$m.

This illustrates the conundrum we face: the X-ray data for
\targetnamew\ imply a colossal bolometric
luminosity,\footnote{Determining the rest-frame 250\,nm--2\,keV
  slope using the observed optical magnitude would yield
  $\alpha_{\rm OX} \approx 1.1$, which would suggest a lower
  bolometric luminosity, yet this slope would then imply heavy UV
  obscuration; thus the approach taken in \S\,\ref{sec:results}
  is physically meaningful and appropriate.} similar in magnitude
to the far-IR luminosity, yet the latter arises from a large disk
and cannot sensibly be powered by the AGN; unlike other dusty
Type-1 AGN, there is no obvious sign of significant excess
rest-frame UV--optical--mid-IR emission from the AGN, yet the
accretion energy must be emerging somewhere, unless our
understanding of how X-ray luminosity maps to bolometric
luminosity is flawed.  The panchromatic SED, low intrinsic X-ray
absorption and general properties of \targetnamew\ therefore
present a considerable puzzle.

A number of plausible scenarios could give rise to the
observed configuration. We deal with three of them in what follows.

\subsection{AGN seen through cavity in an unusually large disk?}

Could a solution to the aforementioned conundrum be related to
power of the AGN and the unusual extent of the dusty starburst in
the galaxy hosting it?  High-resolution spectroscopic imaging of
component W of \targetname\ has revealed molecular gas spread
across a disk that is several times larger ($\approx 3$--7\,kpc
FWHM, depending on the tracer) than the compact, $\approx$kpc
submm continuum emission more typically found in SMGs
\citep[e.g.][]{ikarashi15, simpson15, oteo17hires,
  oteo17almacal2, hodge19, rujopakarn19}, irrespective of the
presence in those SMGs of X-ray-detected AGN
\citep{harrison16}. Such a powerful AGN will rapidly excavate a
central cavity, perhaps with an ionisation cone oriented towards
the NE that can go some way towards explaining the observed
narrow-line properties,\footnote{The velocity offset between the
  narrow lines and the CO is then due to an outflow or winds
  towards the observer.}  perhaps leading to a relatively clear
view of the broad-line region and little X-ray absorption along
the line of sight through the disk\footnote{\citetalias{ivison13}
  and \citet{gomez18} found disk inclinations of $56\pm 10^\circ$
  and $\approx 48^\circ$ respectively, where $0^\circ$ is
  face-on.}  to a central AGN.  It may also be possible that
there is a contribution to the soft X-rays from photoionised gas
and scattering within the ionisation cone, such that the true
X-ray column is larger than that deduced from our simple model
fit.

In this scenario, the observed extinction (see
\S\,\ref{sec:results}) is due to gas and dust local to the AGN,
i.e.\ its obscuring torus in the unified scheme, and/or along the
line of sight through the host galaxy.  $A_V\approx 2.2$
corresponds to over 7~mag of extinction at 125\,nm, which is
roughly the observed $g$ band, such that we would see only the
host galaxy in $g$, consistent with the observed properties and
roughly 5 mag fainter than the prediction for the AGN before
obscuration. We note that the apparent lack of hot dust emission
is a problem for any model that invokes a conventional AGN torus.

If the X-ray opacity proves to be small compared to the dust
attenuation implied by the Balmer decrement\footnote{Measuring
  the Balmer decrement will be relatively straightforward with
  $H$-band spectroscopy.}, this might indicate that the
obscuration arises from a dusty, ionised outflow, akin to that
identified by \citet{mehdipour12}.  Alternatively, if the AGN is
obscured predominantly by dust in the surrounding galaxy then we
would expect a a Seyfert-1-like [O\,{\sc iii}]/H$\alpha$ ratio
and a normal dust-to-gas ratio for the absorber, if we could
combine measurements of X-ray absorption and the broad-line
Balmer decrement.  We could then calculate what fraction of the
AGN power goes into heating the dust in the surrounding galaxy.

\subsection{A second, unseen AGN?}

In the absence of observations that tie down the broad-line
Balmer decrement, or the spatial distribution of [O\,{\sc
  iii}]/H$\beta$, we can speculate that the absence of
significant X-ray absorption --- together with rest-frame optical
features consistent with a Type-1 AGN --- may imply that the
accreting SMBH is not embedded in the gas-rich starburst. This
would, in turn, suggest that energy due to accretion does not
dominate the bolometric luminosity of \targetnamew.  

\citetalias{ivison13} speculated that the mutual proximity and
counter-rotation of the gas disks in components W and T might
explain their unusual luminosities.  Occam's razor might suggest
instead that the near-naked SMBH we observe towards \targetnamew\
--- modulo the lack of associated rest-frame UV--optical emission
--- is associated with another galaxy involved in a merger or
interaction that triggered the starburst
\citep[e.g.][]{ellison19}.

\targetnamew\ would then be expected to contain a second SMBH,
associated with the intense ongoing starburst, plausibly an unseen
Compton-thick AGN. Sadly, although nearby binary AGN are occasionally
revealed \citep[e.g.\ in Mrk~739, at a distance of 130\,Mpc, using
{\it Chandra} --][]{koss11}, the observational challenges associated
with demonstrating the presence of dual or binary AGN at even modest
redshifts are similar or worse than those of identifying distant
interactions and mergers, where the sensitivity and especially spatial
resolution available to X-ray observers are rarely up to the task.

Largely because of SPIRE on the {\it Herschel Space Observatory}
\citep{griffin10, pilbratt10} and the wide-field ground-based
imager, SCUBA-2 \citep{holland13}, many thousands of SMGs are now
known \citep[e.g.][]{eales10, oliver12, geach17}.  While only a
very small fraction of these rest-frame far-IR-selected galaxies
are associated with Type-1 AGN, \citet{knudsen03} describe a
submm-selected, intrinsically hyperluminous quasar,
SMM\,J04135+10277, lensed by the galaxy cluster Abell~478, and
{\it Herschel} led to the detection of many more
\citep[e.g.][]{my15, pitchford16, dongwu16}.  Some others have
Type-2 AGN, or objects believed to be transitioning from Type-2
to Type-1 in the evolutionary scheme proposed by
\cite{sanders88}, e.g.\ the aforementioned SMM\,J02399$-$0136,
with its BAL quasar.  Follow-up submm detections of known
optically-luminous quasars are relatively common \citep[][amongst
others]{isaak94, isaak02, ivison95, omont96qsos, priddey03,
  mainieri05, stacey18, hatziminaoglou18}.  The quasars
BR\,1202$-$0725 and BRI\,1335$-$0417 were amongst the first to
be detected at submm wavelengths, and were later found to have
physically associated SMGs close by \citep[][see also
\citealt{decarli18, decarli19, venemans18}]{omont96, yun00,
  carillibr02, carilli13, salome12, wagg12, wagg14, lu17}, one of
which also harbours an X-ray-luminous AGN \citep{ionochandra06}.

It is conceivable, then, that we have found a system that
contains a HyLIRG, with a buried AGN, close to another more
evolved quasar-like system.  We cannot ignore the stark
difference between the SED of \targetnamew\ and those of other
quasars known to have SMG companions; however, a location in the
outskirts or behind the gas-rich disk of \targetnamew\ may help
explain why the Type-1 AGN is not bright at rest-frame
UV--optical--mid-IR wavelengths.  With sufficient separation,
such a geometry might be revealed using the spatial resolution of
{\it Chandra}, even if it could not be easily distinguished from
our next suggestion.

\subsection{An ejected AGN?}

In another scenario --- discussed in terms of a merging galaxy
pair in the COSMOS field by \citet{civano12} --- asymmetric
emission of gravitational radiation \citep{peres62, bekenstein73}
during the coalescence of two SMBHs with anti-aligned spins
\citep{campanelli07,lz11a,lz11b} and a high mass ratio
\citep{baker08} could have led to the ejection of the newly
formed SMBH from the site of the merger, with a relative velocity
as high as 5,000\,km\,s$^{-1}$.

Such an ejected SMBH is thought unlikely to carry its narrow-line
region along with it \citep{loeb07}, but it could shine for $10^7$\,yr
as a Type-1 AGN, and in an extremely gas-rich environment like that in
\targetnamew\ --- spanning several\footnote{$\approx 7$\,kpc FWHM, as
  measured in CO $J=1$--0 \citepalias{ivison13} so $\approx 10^7$\,yr
  at the maximum plausible velocity, or up to an order of magnitude
  higher at the measured line-of-sight velocity offset betweeen
  H$\alpha$ + [N\,{\sc ii}] and CO.} kpc --- the AGN would give rise
to a narrow-line region as it travels.

We can further speculate that if the SMBH that we observe in
X-rays and in the rest-frame optical has been ejected from the
site of the merger, then the resulting lack of feedback via
powerful AGN-driven winds \citep[e.g.][]{maiolino12, veilleux13,
  veilleux17, cicone14, tombesi15} --- often invoked to regulate
the growth of the stellar spheroidal component of the host galaxy
and the SMBH itself \citep[cf.][cf.\
\citealt{grimmett19}]{ramasawmy19} --- may explain the extreme
nature of the starburst in \targetnamew.  Feedback from
supernovae would be left as the primary regulation mechanism,
perhaps enabling the object to skip quickly to the end of the
aforementioned \citeauthor{sanders88} sequence, from
Compton-thick AGN to naked quasar.  Indeed, recent theoretical
work by \citet{mcalpine19}, based on cosmological hydrodynamical
simulations, suggests that galaxies at $z\approx 2.5$ with high
\lir\ are able to reach and maintain large SFRs because their gas
reservoirs are not depleted by accretion onto their central black
holes, such that their black holes are under-massive.  It would
be interesting if --- in \targetnamew\ --- we have found an
extreme example of this hypothesis, with the absence of a
significantly massive black hole pushing its starburst firmly
into the HyLIRG category.

Although undeniably interesting, the associated implication of
this last scenario --- that SMBHs may permeate intergalactic
space --- is not a topic for this paper.

\section{Summary and concluding remarks}
\label{sec:summary}

We report new X-ray, near-IR and submm observations of the starburst
galaxies that comprise \targetname\ at $z=2.4$.

Our ALMA imaging confirms a more distant, fifth, dust- and gas-rich
member, E, of the \targetname\ protocluster, which is now known to
cover $\approx 15$\,arcsec or $\approx 120$\,kpc.  \targetnamee\ is
extremely red, with an unusually broad CO $J=1$--0 line; it may be a
merger or a colossal disk.

Our {\it XMM-Newton} and KMOS imaging spectroscopy of
\targetnamew\ --- the brightest of the five galaxies, a HyLIRG,
unlensed and extraordinarily luminous, even by the standards of
SMGs --- reveal the presence of an AGN.  \targetnamew\ displays
significant X-ray emission together with bright \Nplus\ lines and
a very broad \ha\ line; the latter implies
$M_{\rm bh}\approx 2\times 10^9$\,M$_\odot$.  For such a dusty
and gas-rich host galaxy, we see surprisingly little intrinsic
absorption towards the AGN,
$N_{\rm H}\approx 5\times 10^{21}$\,cm$^{-2}$, likely with modest
extinction, $A_V\approx 2$.  Our estimate of the bolometric
luminosity of the X-ray-bright AGN is commensurate with the
far-IR luminosity of the starburst, yet we know from spatially
resolved imaging spectroscopy that the system contains a colossal
gas- and dust-rich disk, with no significant temperature
gradient.  Despite the AGN's potential to dominate the overall
power budget, it is therefore not obvious that it does so.

We outline three plausible scenarios that could give rise to the
observed characteristics of \targetnamew, though the lack of
significant rest-frame UV--optical and/or mid-IR emission remains
a puzzle in all of them.

Either we have a relatively clear view of the broad-line region
through the starbursting disk of the host galaxy, with the
powerful AGN having excavated a central cavity, or the AGN is not
embedded in the starburst.  In this second, prosaic option ---
where analogues are known --- we speculate that there are two
SMBHs: one, visible in X-rays, having evolved more quickly
towards the naked quasar phase, in an unseen galaxy or galaxy
remnant that lies very close to the HyLIRG; the second, buried
deep within the dusty starburst, invisible to us.

In our third scenario, the observed SMBH has been ejected from
the region experiencing the starburst, e.g.\ via asymmetric
gravitational radiation during the coalescence of two SMBHs, and
we postulate that the resulting absence of local AGN feedback may
then explain the extreme nature of the starburst.

It is clear that there is considerably more to learn about the role
and impact of the AGN or AGNs in \targetname.  The intrinsic
absorption is detected only barely and is poorly constrained, so it
would be interesting to determine the degree of obscuration through
which we are viewing the AGN.  This could be achieved through a
combination of deeper X-ray data --- to more accurately measure the
absorbing column and to determine whether it is dominated by
reflection and/or has the strong 6.4-keV emission line that usually
characterises this --- and through rest-frame optical spectroscopy to
measure the Balmer decrement of the broad emission lines. Mid-IR
spectroscopy would also be useful to help disentangle the AGN/star
formation contribution to the IR luminosity.

Our findings add to the growing body of evidence that powerful
AGN are ubiquitous amongst HyLIRGs.  However, the nature of the
AGN observed in \targetname\ is not at all what we expected and
we cannot easily reconcile the high bolometric luminosity and
modest intrinsic absorption inferred from our X-ray observations
with the large far-IR-emitting disk and the faint rest-frame
UV--optical--mid-IR portion of its panchromatic SED.

\section*{Acknowledgements}

We acknowledge the many contributions from Iv\'an Oteo, the wisdom of
Alain Omont and Ian Smail, and the generosity of Edith Falgarone,
Hannah Stacey and Song Huang. We are also indebted to the anonymous
referee for an excellent report.

RJI dedicates this paper to his long-time friend and colleague, Wayne
Holland, who passed away in May 2019; without his immensely valuable
contributions to submm instrumentation and astronomical research, we
would still be waiting for ALMA.

This publication makes use of data products from the {\it Wide-field
Infrared Survey Explorer}, which is a joint project of the University
of California, Los Angeles, and the Jet Propulsion
Laboratory/California Institute of Technology, funded by the National
Aeronautics and Space Administration.

This paper makes use of the following ALMA data: 
ADS/JAO.ALMA\#2013.1.00164.S. ALMA is a partnership of ESO
(representing its member states), NSF (USA) and NINS (Japan),
together with NRC (Canada), MOST and ASIAA (Taiwan), and KASI
(Republic of Korea), in cooperation with the Republic of
Chile. The Joint ALMA Observatory is operated by ESO, AUI/NRAO
and NAOJ.



\bibliographystyle{mnras}
\bibliography{rji} 

\bsp 

\label{lastpage}
\end{document}